\documentclass[12pt]{article} 
\usepackage{amssymb} 
\usepackage{amsmath}
\usepackage{amsfonts}

\newcommand{\R}{\mathbb{R}}

\newcommand{\Z}{\mathbb{Z}}

\newcommand{\be}{\begin{equation}}
\newcommand{\bea}{\begin{eqnarray}}
\newcommand{\eea}{\end{eqnarray}}
\newcommand{\nn}{\nonumber}
\newcommand{\kt}{\rangle}
\newcommand{\br}{\langle}

\newcommand{\qq}{{\cal Q}}
\newcommand{\ed}{\end{document}}

\begin{document}

\title{Supersymmetric Dynamical Invariants}
\author{Ali Mostafazadeh\thanks{E-mail address: 
amostafazadeh@ku.edu.tr}\\ \\
Department of Mathematics, Ko\c{c} University,\\
Rumelifeneri Yolu, 80910 Sariyer, Istanbul, Turkey}
\date{ }
\maketitle

\begin{abstract}
We address the problem of identifying the (nonstationary) quantum systems that 
admit supersymmetric dynamical invariants. In particular, we give a general expression 
for the bosonic and fermionic partner Hamiltonians. Due to the supersymmetric nature 
of the dynamical invariant the solutions of the time-dependent Schr\"odinger equation for 
the partner Hamiltonians can be easily mapped to one another. We use this observation to 
obtain a class of exactly solvable time-dependent Schr\"odinger equations. As applications
of our method, we construct classes of exactly solvable time-dependent generalized harmonic 
oscillators and spin Hamiltonians.
\end{abstract}

\baselineskip=24pt

\section{Introduction}

The problem of the solution of the time-dependent Schr\"odinger equation,
	\be
	i\frac{d}{dt}|\psi(t)\kt=H(t)|\psi(t)\kt\;,
	\label{sch-eq}
	\end{equation}
is as old as quantum mechanics. It is well-known that this equation may be reduced to
the time-independent Schr\"odinger equation, i.e., the eigenvalue equation for the 
Hamiltonian, provided that the eigenstates of the Hamiltonian are time-independent.%
\footnote{In this case, the adiabatic approximation is exact, \cite{pra-97a}.} The 
search for exact solutions of the eigenvalue equation for the 
Hamiltonian has been an ongoing effort for the past seven decades. A rather recent 
development in this direction is the application of the ideas of supersymmetric quantum 
mechanics \cite{witten-82}. The main ingredient provided by supersymmetry is that the 
eigenvectors of the bosonic and fermionic partner Hamiltonians are related by a 
supersymmetry transformation \cite{review}. Therefore, one can construct the solutions of 
the eigenvalue problem for one of the partner Hamiltonians, if the other is exactly solvable.
In general, this method cannot be used to relate the solutions of the time-dependent 
Schr\"odinger equation unless the partner Hamiltonians have time-independent eigenvectors.
The aim of this article is to explore the utility of supersymmetry in solving time-dependent 
Schr\"odinger equation for a general class of time-dependent Hamiltonians. 

This problem has been considered by Bagrov and Samsonov \cite{ba-sa} and Cannata {\em 
et al.} \cite{cannata} for the standard Hamiltonians of the form $H=p^2/(2m)+V(x;t)$ 
in one dimension. Our method differs from those of these authors in the following way. 
First, we approach the problem from the point of view of the theory of dynamical invariants 
\cite{lewis-riesenfeld,nova}. Dynamical invariants are certain (time-dependent) operators 
with a complete set of eigenvectors that are exact solutions of the time-dependent 
Schr\"odinger equation. We can easily use the ideas of supersymmetric quantum 
mechanics to relate the solutions of the time-dependent Schr\"odinger equation for two 
different Hamiltonians, if we can identify them with the bosonic and fermionic Hamiltonians 
of a (not necessarily supersymmetric) $\Z_2$-graded quantum system admitting a 
supersymmetric dynamical invariant. Unlike Refs.~\cite{ba-sa} and \cite{cannata},
we consider general even supersymmetric dynamical invariants and use
our recent results on the geometrically equivalent quantum systems \cite{p36} to 
give a complete characterization of the time-dependent Hamiltonians that admit 
supersymmetric dynamical invariants. 

The organization of the article is as follows. In Sections~2, we present a brief review of
the dynamical invariants and survey our recent results on identifying the Hamiltonians that 
admit a given dynamical invariant. In Section~3, we discuss the supersymmetric dynamical
invariants. In section~4, we give a characterization of the quantum systems that admit a 
Hermitian supersymmetric dynamical invariant. In sections~5 and~6, we apply our general results to 
obtain classes of exactly solvable time-dependent generalized harmonic oscillators and spin
systems, respectively . In section~6, we compare our method with that of Refs.~\cite{ba-sa}
and \cite{cannata} and present our concluding remarks.

\section{Dynamical Invariants}

By definition \cite{lewis-riesenfeld,nova}, a dynamical invariant is a nontrivial solution 
$I(t)$ of the Liouville-von-Neumann equation
	\be
	\frac{d}{dt}\,I(t)=i[I(t),H(t)]\;,
	\label{dyn-inv}
	\end{equation}
where $H(t)$ denotes the Hamiltonian. 

Consider a Hermitian Hamiltonian $H(t)$ admitting a Hermitian dynamical invariant $I(t)$, 
and suppose that $I(t)$ has a discrete spectrum\footnote{The generalization to a continuous 
spectrum is not difficult.}. Then, Eq.~(\ref{dyn-inv}) may be used to show that the 
eigenvalues $\lambda_n$ of $I(t)$ are constant and the eigenvectors $|\lambda_n,a;t\kt$ 
yield the evolution operator $U(t)$ for the Hamiltonian $H(t)$ according to
	\be
	U(t)=\sum_n\sum_{a=1}^{d_n} u^n_{ab}(t)|\lambda_n,a;t\kt\br\lambda_n,b;0|\;.
	\label{U}
	\end{equation}
Here $n$ is a spectral label, $a\in\{1,2,\cdots,d_n\}$ is a degeneracy label, $d_n$ is the 
degree of degeneracy of $\lambda_n$, the eigenvectors  $|\lambda_n,a;t\kt$ are assumed to 
form a complete orthonormal basis of the Hilbert space, $u^n_{ab}(t)$ are the entries of the 
solution of the matrix Schr\"odinger equation:
	\bea
	i\frac{d}{dt}\,u^n(t)&=&\Delta(t)u^n(t)\,,~~~~u^n(0)=1\,,
	\label{u-n}\\
	\Delta(t)&:=&{\cal E}^n(t)-{\cal A}^n(t)\;,
	\label{Delta}
	\eea
and ${\cal E}^n(t)$ and ${\cal A}^n(t)$ are matrices with entries
	\be
	{\cal E}^n_{ab}:=\br\lambda_n,a;t|H(t)|\lambda_n,b;t\kt\,,~~~~
	{\cal A}^n_{ab}:=i\br\lambda_n,a;t|\frac{d}{dt}|\lambda_n,b;t\kt,
	\label{E-A}
	\end{equation}
respectively, \cite{jpa98,nova}. Note that ${\cal E}^n(t),{\cal A}^n(t),\Delta^n(t)$ are 
Hermitian matrices and $u^n(t)$ is unitary.

In view of Eq.~(\ref{U}), 
	\be
	|\psi_n,a;t\kt:=U(t)|\lambda_n,a;0\kt=
	\sum_{b=1}^{d_n} u^n_{ba}(t)|\lambda_n,b;t\kt
	\label{solution}
	\end{equation}
are solutions of the Schr\"odinger equation~(\ref{sch-eq}). These solutions actually form
a complete orthonormal set of eigenvectors of $I(t)$. We may use this observation or
alternatively Eq.~(\ref{U}) to show
	\be
	I(t)=U(t)I(0)U^\dagger(t)\;.
	\label{I=UIU}
	\end{equation}

Now, suppose that $I(t)$ is obtained from a parameter-dependent operator $I[\bar R]$ as 
$I(t)=I[\bar R(t)]$ where 
	\begin{itemize}
	\item[1.] $\bar R=(\bar R^1,\bar R^2,\cdots,\bar R^r)$, $\bar R^i$ are real 
	parameters denoting the coordinates of points of a parameter manifold $\bar M$;
	\item[2.] $\bar R(t)$ determines a smooth curve in $\bar M$;
	\item[3.] $I[\bar R]$ is a Hermitian operator with a discrete spectrum;
	\item[4.] (in local coordinate patches of $\bar M$) the eigenvectors 
	$|\lambda_n,a;\bar R\kt$ of $I[\bar R]$, i.e., the solutions of
		\be
		I[\bar R]|\lambda_n,a;\bar R\kt=\lambda_n|\lambda_n,a;\bar R\kt
		~~~~{\rm with}~~~~a\in\{1,2,\cdots,d_n\}\;,
		\label{eg-va}
		\end{equation}
	are smooth (single-valued) functions of $\bar R$;
	\item[5.] $\lambda_n$ and $d_n$ are independent of $\bar R$;
	\item[6.] $|\lambda_n,a;\bar R\kt$ form a complete orthonormal basis.
	\end{itemize}
In the following, we shall identify $|\lambda_n,a;t\kt$ with $|\lambda_n,a;\bar R(t)\kt$ and 
express $|\lambda_n,a;\bar R\kt$ in the form
	\be
	|\lambda_n,a;\bar R\kt=W[\bar R]|\lambda_n,a;\bar R(0)\kt,
	\label{L=WL}
	\end{equation}
where $W[\bar R]$ is a unitary operator and $W=W[\bar R]$ defines a single-valued 
function of $\bar R$. Eqs.~(\ref{eg-va}) and (\ref{L=WL}) suggest
	\be
	I[\bar R]=W[\bar R]I(0)W[\bar R]^\dagger\;.
	\label{I=WIW}
	\end{equation}

For a closed curve $\bar R(t)$, there exists $T\in\R^+$ such that $\bar R(T)=
\bar R(0)$, and the quantity
	\be
	\Gamma^n(T):={\cal T}e^{i\int_0^T {\cal A}^n(t')dt'}={\cal P}e^{i\oint A^n}
	\label{G}
	\end{equation}
yields the non-Abelian cyclic geometric phase \cite{anandan-88,gp,nova} associated with the 
solution~$|\psi_n;a;t\kt$. In Eq.~(\ref{G}), ${\cal T}$ and ${\cal P}$ respectively denote 
the time-ordering and path-ordering operators, the loop integral is over the closed path
$\bar R(t)$, and $A^n$ is the nondegenerate non-Abelian generalization of the Berry 
connection one-form \cite{berry-84,simon-83}. The latter is defined in terms of its
matrix elements:
	\be
	A^n_{ab}[\bar R]:=i\br\lambda_n,a;\bar R|\bar d|\lambda_n,b;\bar R\kt\;,
	\label{connection}
	\end{equation}
where $\bar d=\sum_{i}d{\bar R}^i\,\partial/\partial{\bar R}^i$ is the exterior derivative 
operator on $\bar M$. If $\lambda_n$ is nondegenerate, $\Gamma^n(t)$ is just a phase factor.
It coincides with the (nonadiabatic) geometric phase of Aharonov and Anandan
\cite{aa}.

Next, we introduce $W(t):=W[\bar R(t)]$. Then as discussed in \cite{p36}, Hermitian Hamiltonians
that admit the invariant 
	\be
	I(t)=W(t)I(0)W(t)^\dagger
	\label{I=WIW-t}
	\end{equation}
have the form
	\be
	H(t)=W(t)Y(t)W(t)^\dagger-iW(t)\frac{d}{dt}\,W(t)^\dagger\;,
	\label{H=}
	\end{equation}
where $Y(t)$ is any Hermitian operator commuting with $I(0)$. Note that according to
Eq.~(\ref{H=}), $H(t)$ is related to $Y(t)$ by a time-dependent (canonical) unitary 
transformation of the Hilbert space \cite{pla-97,jmp-97b,p36,nova}, namely $|\psi(t)\kt\to
W(t)|\psi(t)\kt$. This observation may be used to express the evolution operator $U(t)$ of
$H(t)$ in the form
	\be
	U(t)=W(t) V(t)\;,
	\label{U=WE}
	\end{equation}
where $V(t):={\cal T}e^{-i\int_0^t Y(t')dt'}$ is the evolution operator for $Y(t)$.
Note that $Y(t)$ commutes with $I(0)$, therefore if $I(0)$ has a nondegenerate spectrum, $Y(t)$ has a constant
eigenbasis. In this case, $Y(t)$ with different $t$ commute and
	\be
	V(t)= e^{-i\int_0^t Y(t')dt'}.
	\label{V=E}
	\end{equation}
Having expressed $U(t)$ in terms of $Y(t)$ and $W(t)$, we can write the 
solutions~(\ref{solution}) of the Schr\"odinger equation in the form
	\be
	|\psi_n,a;t\kt:=W(t)V(t) |\lambda_n,a;0\kt =\sum_{b=1}^{d_n}V^n_{ab}(t)W(t)
	|\lambda_n,b;0\kt=\sum_{b=1}^{d_n}V^n_{ab}(t)|\lambda_n,b;t\kt\;,
	\label{solution-y}
	\end{equation}
where $V^n_{ab}(t):=\br\lambda_n,a;0|V(t)|\lambda_n,b;0\kt$. If $Y(t)$ with different values of $t$ commute, this equation
takes the form
	\be
	|\psi_n,a;t\kt:=e^{-i\int_0^t y^n_a(t')dt'}|\lambda_n,a;t\kt\;,
	\nn
	\end{equation}
where $y^n_a(t):=\br\lambda_n,a;0|Y(t)|\lambda_n,a;0\kt$.

We conclude this section by emphasizing that Eqs.~(\ref{H=}) and (\ref{U=WE}) are valid
for any time-dependent unitary operator $W(t)$ satisfying Eq.~(\ref{I=WIW-t}). For example,
one may identify $W(t)$ with the evolution operator of another Hamiltonian that admits the 
same invariant $I(t)$. Note that in general such a choice of $W(t)$ cannot be expressed as 
the image of a curve $\bar R(t)$ under a single-valued function $W[\bar R]$. In particular, 
$|\lambda_n,a;t\kt':=W(t)|\lambda_n;a;0\kt$ cannot be written as $|\lambda_n,a;\bar R(t)\kt$
for parameter-dependent vectors $|\lambda_n,a;\bar R\kt$ that are single-valued functions
of $\bar R$. This in turn implies that $|\lambda_n,a;t\kt'$ cannot be used in the calculation
of the geometric phases.

\section{$\Z_2$-Graded Systems Admitting Supersymmetric Invariants}

A $\Z_2$-graded quantum system \cite{p34b} is a system whose Hilbert space 
$\tilde{\cal H}$ is the direct sum of two of its nontrivial subspaces ${\cal H}_\pm$, i.e., 
$\tilde{\cal H}={\cal H}_+\oplus{\cal H}_-$, and whose Hamiltonian maps 
${\cal H}_\pm$ to ${\cal H}_\pm$. The elements of ${\cal H}_+$ and ${\cal H}_-$ are 
respectively called bosonic and fermionic state vectors, or graded state vectors with definite 
grading (or chirality) 0 and 1. Operators preserving the grading of the graded state vectors 
are called even operators. Those that change the grading of these state vectors are called odd
operators.

In the two-component representation of the Hilbert space, where the first component 
$|\psi_+\kt$ denotes the bosonic and the second component $|\psi_-\kt$ denotes the 
fermionic part of a state vector $|\psi\kt=|\psi_+\kt+|\psi_-\kt$, the Hamiltonian has the form
	\be
	H(t)=\left(\begin{array}{cc}
	H_+(t) & 0\\
	0 & H_-(t) \end{array}\right).
	\label{H=HH}
	\end{equation}
Here $H_+(t):{\cal H}_+\to{\cal H}_+$ and $H_-(t):{\cal H}_-\to{\cal H}_-$ are 
Hermitian operators. They are respectively called the bosonic and fermionic Hamiltonians.

Now, suppose that ${\cal H}_+={\cal H}_-=:{\cal H}$ and consider a parameter-dependent 
odd operator $\qq=\qq[\bar R]$ and an even Hermitian operator $I=I[\bar R]$ that satisfy 
the algebra of $N=1$ supersymmetric quantum mechanics \cite{review}:
	\be
	\qq^2=0,~~~~[\qq,I]=0,~~~~\{\qq,\qq^\dagger\}=2I.
	\label{susy}
	\end{equation}
In particular, suppose that in the two-component representation of the Hilbert space, 
	\[\qq= \left(\begin{array}{cc}
	0 & 0\\
	d & 0 \end{array}\right),\]
where $d=d[\bar R]$ is a linear operator. This choice of $\qq$ 
satisfies the superalgebra~(\ref{susy}) provided that
	\be
	I=\left(\begin{array}{cc}
	I_+ & 0\\
	0 & I_- \end{array}\right),
	\label{I=II}
	\end{equation}
with
	\be
	I_+:= \frac{1}{2}\, d^\dagger d \,,~~~~
	I_-:= \frac{1}{2}\, d d ^\dagger\,.
	\label{I-I}
	\end{equation}

As is well-known from the study of the spectral properties of supersymmetric systems, one 
can use Eqs.~(\ref{I-I}) to derive the following properties of $I_\pm$.
	\begin{itemize}
	\item[-] $I_+$ and $I_-$ have nonnegative spectra with the same set of positive 
	eigenvalues $\lambda_n$;
	\item[-] The degree of degeneracy $d_n$ of $\lambda_n>0$ as an eigenvalue of 
	$I_+$ is the same as its degree of degeneracy as an eigenvalue of $I_-$;
	\item[-] Orthonormal eigenvectors $|\lambda_n,a,\pm;\bar R\kt$ of $I_\pm[\bar R]$
 	associated with $\lambda_n>0$ are related according to
	\bea
	d[\bar R] |\lambda_n,a,+;\bar R\kt&=&\sqrt{2\lambda_n}\sum_{b=1}^{d_n}
	v_{ba}[\bar R]|\lambda_n,b,-;\bar R\kt \;,
	\label{dL=L-1}\\
	d^\dagger[\bar R] |\lambda_n,b,-;\bar R\kt&=&\sqrt{2\lambda_n}\sum_{a=1}^{d_n}
	v_{ab}[\bar R]^\dagger|\lambda_n,a,+;\bar R\kt \;,
	\label{dL=L-2}
	\eea
	where $v_{ab}[\bar R]$ are the entries of a unitary $d_n\times d_n$ matrix 
	$v[\bar R]$. In particular, for a given orthonormal set 
	$\{|\lambda_n,a,+;\bar R\kt ~|~\lambda_n>0\}$ of the eigenvectors of $I_+[\bar R]$, 
		\[|\lambda_n,a,-;\bar R\kt:=(2\lambda_n)^{-1/2}d[\bar R]|\lambda_n,a,+;\bar R\kt\]
	form a complete orthonormal eigenbasis of $I_-[\bar R]$ for 
	${\cal H}_--{\rm Ker}(I_-[\bar R])$. Here `Ker' denotes the kernel or the 
	eigenspace with zero eigenvalue.
	\end{itemize}

Next, we introduce $I(t):=I[\bar R(t)]$ and $I_\pm(t):=I_\pm[\bar R(t)]$ for some curve
$\bar R(t)$ in the parameter space $\bar M$ and demand that $I(t)$ is a dynamical invariant 
for the Hamiltonian $H(t)$. In view of Eqs.~(\ref{dyn-inv}), (\ref{H=HH}), (\ref{I=II}), 
and (\ref{I-I}), $I_\pm(t)$ is a dynamical invariant for $H_\pm(t)$. 

We can write $I_\pm[\bar R]$ in the form~(\ref{I=WIW}) by requiring  $d[\bar R]$ to 
satisfy 
	\be
	d[\bar R]= W_-[\bar R]d(0)W_+[\bar R]^\dagger,
	\label{d=wdw}
	\end{equation}
where $d(t):=d[\bar R(t)]$ and $W_\pm[\bar R]$ fulfil
	\be
	|\lambda_n,a,\pm;\bar R\kt=W_\pm[\bar R]|\lambda_n,a,\pm;\bar R(0)\kt.
	\label{L=WL-2}
	\end{equation}
In view of Eqs.~(\ref{I-I}) and~(\ref{d=wdw}),
	\be
	I_\pm[\bar R]=W_\pm[\bar R] I_\pm(0) W_\pm[\bar R]^\dagger\;.
	\label{I=WIW-2}
	\end{equation}

Moreover, employing Eqs.~(\ref{H=}) and (\ref{U=WE}), we can express the Hamiltonians 
$H_\pm(t)$ and their evolution operators $U_\pm(t)$ in the form
	\bea
	H_\pm(t)&=&W_\pm(t)Y_\pm(t)W_\pm(t)^\dagger
	-iW_\pm(t)\frac{d}{dt}W_\pm(t)^\dagger,
	\label{H-pm=}\\
	U_\pm(t)&=&W_\pm(t) V_\pm(t),~~~~
	V_\pm(t):={\cal T}\,e^{-i\int_0^t Y_\pm(t')dt'},
	\label{U-pm=}
	\eea
where $Y_\pm(t)$ are Hermitian operators satisfying
	\be
	[Y_\pm(t),I_\pm(0)]=0\;.
	\label{Y-I=zero}
	\end{equation}
For example, we can choose $Y_\pm(t)=P_t(I_\pm(0))$ where $P_t$ is a polynomial with
time-dependent coefficients. 

In view of Eq.~(\ref{solution-y}), we have the following set of orthonormal solutions of
the Schr\"odinger equation for the Hamiltonians $H_\pm(t)$.
	\be
	|\psi_n,a,\pm;t\kt =\sum_{b=1}^{d_n}V^n_{ab\pm}(t)|\lambda_n,b,\pm;t\kt\;,
	\label{solution-pm}
	\end{equation}
where $V^n_{ab\pm}(t):=\br\lambda_n,a,\pm;0|V_\pm(t)|\lambda_n,b,\pm;0\kt$. As we 
discussed above, given an eigenbasis $|\lambda_n,a,+;t\kt$ for $I_+(t)$ we can set
	\[|\lambda_n,a,-;t\kt=(2\lambda_n)^{-1/2}d(t)|\lambda_n,a,+;t\kt~~~{\rm for}~~~ \lambda_n>0.\] 
This identification may be used to relate the solutions~(\ref{solution-pm}) according to
	\be
	|\psi_n,a,-;t\kt =(2\lambda_n)^{-1/2}\sum_{b,c=1}^{d_n} V^{n*}_{ca+}(t)V^n_{cb-}(t)
	d(t)|\psi_n,b,+;t\kt\,,~~~~\forall\lambda_n>0\,.
	\label{psi=psi}
	\end{equation}
For the case where $Y_\pm(t)$ with different $t$ commute, Eqs.~(\ref{solution-pm}) and (\ref{psi=psi}) take the form
	\bea
	|\psi_n,a,\pm;t\kt&=&e^{-i\int_0^t y^n_{a\pm}(t')dt'} |\lambda_n,a,\pm;t\kt\;,
	\label{solution-pm-1}\\
	|\psi_n,a,-;t\kt&=&(2\lambda_n)^{-1/2}e^{i\int_0^t [y^n_{a+}(t')-y^n_{a-}(t')]dt'} 
				d(t)|\psi_n,a,+;t\kt\,,~~~~\forall\lambda_n>0\,,
	\label{psi=psi-1}
	\eea
respectively. Here $y^n_{a\pm}(t):=\br\lambda_n,a,\pm;0|Y_\pm(t)|\lambda_n,a,\pm;0\kt$. 

The above construction is valid for any choice of time-dependent unitary operators $W_\pm(t)$ 
satisfying
	\be
	d(t)=W_-(t)d(0)W_+(t)^\dagger.
	\label{d=WdW-t}
	\end{equation}
These observations together with Eq.~(\ref{I=UIU})  suggests a method of generating a 
class of exactly solvable time-dependent Schr\"odinger equations. This is done according to 
the following prescription.
	\begin{itemize}
	\item[1.] Choose a Hamiltonian $H_+(t)$ whose  time-dependent Schr\"odinger 	
	equation is exactly solvable, i.e., its evolution operator $U_+(t)$ is known;
	\item[2.] Choose an arbitrary constant operator $d_0$ and a unitary operator 
	$W_-(t)$ satisfying $W_-(0)=1$;
	\item[3.] Set $I_+(0):=d_0^\dagger d_0/2$, $I_-(0):=d_0d_0^\dagger/2$, and 
	$W_+(t)=U_+(t)$. Then, by construction $I_+(t):=U_+(t)I_+(0)U_+^\dagger$ is a
	dynamical invariant for $H_+(t)$. It also satisfies $I_+(t)=d(t)^\dagger d(t)/2$ for
	$d(t):=W_-(t)d_0U_+(t)^\dagger$. Note that, in view of Eq.~(\ref{H-pm=}) and the 
	Schr\"odinger equation
		\be 	
		i\frac{d}{dt}\,U_+(t)=H_+(t)U_+(t)\;,
		\label{sch-eq-U}
		\end{equation}
	this choice	of $W_+(t)$ corresponds to taking $Y_+(t)=0$.
	\item[4.] Let $Y_-(t)$ be a Hermitian operator commuting with $I_-(0)$. Then
	according to Eq.~(\ref{H-pm=}), $H_+(t)$ and
		\be
		H_-(t):=W_-(t)Y_-(t)W_-(t)^\dagger-iW_-(t)\frac{d}{dt}\,W_-(t)^\dagger
		\label{H-minus}
		\end{equation}
	are partner Hamiltonians, and $H_-(t)$ admits the invariant $I_-(t):=W_-(t)I_-(0)
	W_-(t)^\dagger$. 
	\end{itemize}

The choice $W_+(t)=U_+(t)$ also implies that $|\lambda_n,a,+;t\kt =U_+(t)
|\lambda_n,a,+;0\kt$ are solutions of the Schr\"odinger equation~(\ref{sch-eq}) for $H_+(t)$. 
Furthermore, for all $\lambda_n>0$,
	\be
	|\psi_n,a,-;t\kt =(2\lambda_n)^{-1/2}\sum_{b=1}^{d_n} V^n_{ab-}(t)d(t)|\lambda_n,b,+;t\kt
	\label{psi=psi-2}
	\end{equation}
are solutions of the Schr\"odinger equation for the Hamiltonian $H_-(t)$. These 
solutions span $ {\cal H}_--{\rm Ker}(I_-(0)) ={\cal H}_--{\rm Ker}(d_0)$. Again if $Y_-(t)$ with different values
of $t$ commute, we have
	\be
	|\psi_n,a,-;t\kt =(2\lambda_n)^{-1/2}e^{-i\int_0^t y^n_{a-}(t')dt'} 
	d(t)|\lambda_n,a,+;t\kt,~~~\forall\lambda_n>0.
	\label{psi=psi-2-1}
	\end{equation}

One can also employ an alternative construction for the Hamiltonian $H_-(t)$ in which one
still defines $I_+(t)$ according to $I_+(t):=U_+(t)I_+(0)U_+(t)^\dagger$ but uses another unitary
operator $W_+(t)$ to express it as $I_+(t)=W_+(t)I_+(0)W_+(t)^\dagger$. In this way, one may
choose $W_+(t)$ to be the image of a curve $\bar R(t)$ in a parameter space $\bar M$ under a 
single-valued function $W_+=W_+[\bar R]$. This is especially convenient for addressing 
the geometric phase problem for the Hamiltonians $H_\pm(t)$. Following this approach, 
one must determine $Y_+(t)$ according to Eq.~(\ref{H-pm=}), i.e.,
	\be
	Y_+(t)=W_+(t)^\dagger H_+(t)W_+(t)-iW_+(t)^\dagger\frac{d}{dt}\,W_+(t).
	\label{Y=}
	\end{equation}
One then obtains the Hamiltonian $H_-(t)$ by substituting (\ref{Y=}) in (\ref{H-pm=}).

\section{Partner Hamiltonians for the Unit Simple Harmonic Oscillator Hamiltonian}

In this section we explore the partner Hamiltonians for the Hamiltonian of the unit simple harmonic 
oscillator:
	\be
	H_+=\frac{1}{2}\,(p^2+x^2)=a^\dagger a+\frac{1}{2}=
	a\,a^\dagger-\frac{1}{2}\;.
	\label{sho}
	\end{equation}
Here $p$ and $x$ are respectively the momentum and position operators and 
$a:=(x+ip)/\sqrt{2}$. Let $W_-(t)$ be a unitary operator satisfying $W_-(0)=1$ and
	\be
	d(t):=W_-(t)a^\dagger\;.
	\label{d=wa}
	\end{equation}
Then $I_+=d^\dagger d/2=a\,a^\dagger/2$ is a dynamical invariant for $H_+$. This invariant together with 	
	\be
	I_-(t)=\frac{1}{2}\,d(t)d(t)^\dagger=\frac{1}{2}W_-(t)a^\dagger a\,W_-(t)^\dagger
	\label{I-}
	\end{equation}
form a supersymmetric dynamical invariant. The associated `fermionic' partner Hamiltonian
is given by Eq.~(\ref{H-minus}) where $Y_-(t)$ is a Hermitian operator commuting with
$I_-(0)= a^\dagger a/2$.

For example, let 
	\be
	Y_-(t)= \frac{f(t)}{4}\,(2a^\dagger a+1)= \frac{f(t)}{4}\,(p^2+x^2)	\;,
	\label{Y-=}
	\end{equation}
where $f(t)$ is a real-valued function, and $W_-(t)=W_-[\theta(t),\varphi(t)]$ where
	\bea
	W_-[\theta,\varphi]&:=& e^{-i\varphi K_3}e^{-i\theta K_2}e^{i\varphi K_3}\;,
	\label{su11-1}\\
	K_1&:=&\frac{1}{4}\,(x^2-p^2),~~~
	K_2:=-\frac{1}{4}\,(xp+px),~~~
	K_3:=\frac{1}{4}\,(x^2+p^2),
	\label{su11}
	\eea
$\theta\in\R$, and $\varphi\in[0,2\pi)$. Note that $Y_-(t)$ with different values of $t$ commute and
the operators $K_i$ are generators of the group $SU(1,1)$ in its oscillator representation. The
parameter space of the operator $W_-$ is the unit hyperboloid:
	\[\bar M=\{(\bar R^1,\bar R^2,\bar R^3)\in\R^3~|~-(\bar R^1)^2-(\bar R^2)^2+
	(\bar R^3)^2=1\}.\]

We have made the choices~(\ref{Y-=}) and (\ref{su11-1}) for $Y_-(t)$ and $W_-(t)$ in view of the
following considerations.
	\begin{itemize}
	\item[1.] Up to a trivial addition of a multiple of identity, (\ref{Y-=}) is the most general
	expression for a second order differential operator commuting with $I_-(0)$.
	\item[2.] Every element of (the oscillator representation of the) Lie algebra of $SU(1,1)$ 
	may be expressed as $W_-Y_-W_-^\dagger$ with $Y_-$ and $W_-$ given by 
	Eqs.~(\ref{Y-=}) and (\ref{su11-1}), respectively. In particular, as we show in the following, 	
	these choices lead to the most general expression for an invariant $I_-(t)$ and a Hamiltonian
	$H_-(t)$ belonging to (the oscillator representation of the) Lie algebra of $SU(1,1)$. 
	\end{itemize}

In order to compute the Hamiltonian $H_-(t)$, we substitute Eqs.~(\ref{Y-=}) and (\ref{su11-1}) in 
Eq.~(\ref{H-minus}) and use the $su(1,1)=so(2,1)$ algebra,
	\[ [K_1,K_2]=-iK_3,~~~[K_2,K_3]=iK_1,~~~[K_3,K_1]=iK_2,\]
and the Backer-Campbell-Hausdorff formula to compute the right-hand side of the resulting
equation. We then find, after a rather lengthy calculation,
	\be
	H_-(t)=\sum_{i=1}^3 R^i(t) K_i,
	\label{gho}
	\end{equation}
where
	\bea
	R^1(t)&:=&\sinh\theta(t)\cos\varphi(t)[2f(t)-\dot\varphi(t)]-\sin\varphi(t)\dot\theta(t),
	\label{R1}\\
	R^2(t)&:=&\sinh\theta(t)\sin\varphi(t)[2f(t)-\dot\varphi(t)]+\cos\varphi(t)\dot\theta(t),
	\label{R2}\\
	R^3(t)&:=&2\cosh\theta(t)f(t)+[1-\cosh\theta(t)]\dot\theta(t),
	\label{R3}
	\eea
and a dot denotes a time derivative.

As seen from Eqs.~(\ref{su11}) and (\ref{gho}), $H_-(t)$ is the Hamiltonian of a 
time-dependent generalized harmonic oscillator \cite{nova} with three free functions $f(t)$,
$\theta(t)$, and $\varphi(t)$. According to our general analysis, the corresponding 
Schr\"odinger equation is exactly solvable. The evolution operator is given by
	\[U_-(t)= e^{-i\varphi(t) K_3}e^{-i\theta(t) K_2}e^{iK_3[\varphi(t)-F(t)]},\]
where $F(t)=\int_0^t f(t')dt'$. Furthermore, we can use the stationary solutions of the
Schr\"odinger equation for the unit simple harmonic oscillator (\ref{sho}) to construct
solutions of the Schr\"odinger equation for $H_-(t)$. The stationary solutions for the
Hamiltonian~(\ref{sho}) are
	\be
	|\psi_n,+;t\kt:=e^{-itE_n} |n\kt,
	\label{psi-n=5}
	\end{equation}
where $E_n=n+1/2$, $|n\kt=(n!)^{-1/2}a^{\dagger n}|0\kt$, and $|0\kt$ is the ground
state vector for the unit simple harmonic oscillator (\ref{sho}) given by $\br x|0\kt=
\pi^{-1/4}e^{-x^2/2}$. In view of Eqs.~(\ref{psi=psi-1}), (\ref{d=wa}), and (\ref{psi-n=5}), 
we have the following orthonormal solutions of the Schr\"odinger equation for $H_-(t)$.
	\be	
	|\psi_n,-;t\kt=(n+1)^{-1/2}e^{-itE_n} 
	e^{-i F(t) K_3}W_-(t)a^\dagger |n\kt\
	=e^{-i\zeta_n(t)} e^{-i[F(t)+\varphi(t)]K_3}e^{-i\theta(t) K_2}|n+1\kt\,,
	\label{psi=psi-5}
	\end{equation}
where $\zeta_n(t):=[t-\varphi(t)/2]n+t/2-3\varphi(t)/4$. Next, we use the identity
\cite{jpa-88a}
	\[ e^{i\theta K_2}|x\kt=|e^{\theta/2}x\kt,\]
and the expression for the propagator of the unit simple harmonic oscillator
\cite{holstein}, namely
	\[ U(x,t;x',0):=\br x|U(t)|x'\kt=(2\pi i\sin t)^{-1/2}
	e^{\frac{i[(x^2+{x'}^{2})\cos t-2xx']}{2\sin t}}\;,\]	
to compute the solutions~(\ref{psi=psi-5}) in the position representation. This yields
	\[\br x|\psi_n,-;t\kt=e^{-i\zeta_n(t)}\int_{-\infty}^\infty 
	U(x,F(t)+\varphi(t);x',0)\phi_{n+1}(e^{\theta(t)/2}x')dx',\]
where $\phi_n(x):=\br x|n\kt$ are the eigenfunctions of the unit simple harmonic 
oscillator Hamiltonian~(\ref{sho}).

\section{Partner Hamiltonians for the Dipole Interaction Hamiltonian of a Spinning Particle in a
Constant Magnetic Field}

Consider the dipole interaction Hamiltonian of a spinning particle in a constant magnetic field:
	\be
	H=b J_3\;,
	\label{spin-const}
	\end{equation}
where $b$ is constant (-Larmor frequency), the magnetic field is assumed to be directed along
the $z$-direction, and $J_3$ denotes the $z$-component of the angular momentum operator 
${\bf J}=(J_1,J_2,J_3)$ of the particle. Let $W_-(t)$ be a unitary operator satisfying $W_-(0)=1$
and
	\be
	d(t):=W_-(t)J_+\;,
	\label{d=wa-spin}
	\end{equation}
where $J_\pm:=J_1\pm i J_2=J_\mp^\dagger$. Then, in view of the identity
	\be
	[J_-J_+,J_3]=0\;,
	\label{identity-spin}
	\end{equation}
the operator
	\be
	I_+=d^\dagger d/2=J_-J_+/2
	\label{I+-spin}
	\end{equation}
is a dynamical invariant for $H_+$. Eq.~(\ref{identity-spin}) follows from the $su(2)=so(3)$ algebra,
	\be
	[J_1,J_2]=iJ_3,~~~[J_2,J_3]=iJ_1,~~~[J_3,J_1]=iJ_2,
	\label{su2-algebra}
	\end{equation}
satisfied by $J_i$, the fact that ${\bf J}^2$ is a Casimir operator, i.e., $[{\bf J}^2,J_i]=0$, and the 
relation
	\be
	J_-J_+= J_1^2+J_2^2-J_3 = {\bf J}^2-J_3(J_3+1)\,.
	\label{-+=}
	\end{equation}

The invariant $I_+$ together with 
	\be
	I_-(t)=\frac{1}{2}\,d(t)d(t)^\dagger=\frac{1}{2}W_-(t)J_+J_-W_-(t)^\dagger
	\label{I--spin}
	\end{equation}
form a supersymmetric dynamical invariant. The associated `fermionic' partner Hamiltonian
is given by Eq.~(\ref{H-minus}) where $Y_-(t)$ is a Hermitian operator commuting with
$I_-(0)=J_+J_-/2$.

Next, we note that $[J_+J_-,J_3]=0$. This suggests that we may choose $Y_-(t)$ as a polynomial
in $J_3$ with time-dependent coefficients. For example, we may set
	\be
	Y_-(t)=f(t) J_3
	\label{Y-=spin}
	\end{equation}
where $f(t)$ is a real-valued function. With this choice of $Y_-$, we can construct a class of 
partner Hamiltonians $H_-(t)$ for $H_+$ representing the dipole interaction of a spinning
particle in a time-dependent magnetic field, provided that we choose $W_-(t)
=W_-[\theta(t),\varphi(t)]$ according to \cite{bohm-qm,nova}
	\be
	W_-[\theta,\varphi]:= e^{-i\varphi J_3}e^{-i\theta J_2}e^{i\varphi J_3}\;,
	\label{su2}
	\end{equation}
where $\theta\in[0,\pi)$ and $\varphi\in[0,2\pi)$. Note that again $Y_-(t)$ with different values of $t$
commute, the parameter space of the operator $W_-$ is the unit sphere, and $\theta$ and $\varphi$ are respectively the 
polar and azimuthal angles.\footnote{As described in Refs.~\cite{bohm-qm,nova}, it turns out that $W_-$ as given by
Eq.~(\ref{su2}) fails to be single-valued at the south pole ($\theta=\pi$). One can alternatively change the
sign of $\theta$ on the right-hand side of (\ref{su2}), in which case $W_-$ becomes single-valued
for all values of $\theta$ and $\varphi$ except for $\theta=0$, i.e., the north pole.}

The calculation of Hamiltonian $H_-(t)$ for these choices of $Y_-$ and $W_-$ is similar to that of
section~4. Substituting Eqs.~(\ref{Y-=spin}) and (\ref{su2}) in Eq.~(\ref{H-minus}) and using the 
$su(2)$ algebra (\ref{su2-algebra}) and the Backer-Campbell-Hausdorff formula, we find 
	\be
	H_-(t)=\sum_{i=1}^3 R^i(t) J_i,
	\label{spin}
	\end{equation}
where
	\bea
	R^1(t)&:=&\sin\theta(t)\cos\varphi(t)[2f(t)-\dot\varphi(t)]-\sin\varphi(t)\dot\theta(t),
	\label{R1-spin}\\
	R^2(t)&:=&\sin\theta(t)\sin\varphi(t)[2f(t)-\dot\varphi(t)]+\cos\varphi(t)\dot\theta(t),
	\label{R2-spin}\\
	R^3(t)&:=&2\cos\theta(t)f(t)+[1-\cos\theta(t)]\dot\theta(t),
	\label{R3-spin}
	\eea
As seen from these equations, the fermionic partner Hamiltonians~(\ref{spin}) to the bosonic
Hamiltonian~(\ref{spin-const}) also belong to the Lie algebra $su(2)$; they form a three-parameter
family of dipole Hamiltonians describing spinning particles in time-dependent magnetic fields. The
solution of the Schr\"odinger equation for this type of Hamiltonians has been extensively studied in the 
literature. A rather comprehensive list of references may be found in \cite{nova}. 

In view of Eqs.~(\ref{U-pm=}), (\ref{Y-=spin}), and (\ref{su2}), the evolution operator for the 
Hamiltonian~(\ref{spin}), for arbitrary choices of functions $f,\theta,$ and $\varphi$, is given by
	\be
	U_-(t)=e^{-i\varphi(t)J_3}e^{-i\theta J_2}e^{i[\varphi(t)-F(t)]J_3}\;,
	\label{U--spin}
	\end{equation}
where $F(t):=\int_0^t f(t')dt'$. Moreover, using the supersymmetric nature of our construction, we
may construct a set of orthonormal solutions $|\psi_m,-;t\kt$ of the Schr\"odinger equation 
for this Hamiltonian from those of the constant Hamiltonian (\ref{spin-const}). 

In order to compute these solutions, we first note that $W_+(t)=1$. Therefore, in view of
Eq.~(\ref{Y=}), $Y_+(t)=H_+(t)=bJ_3$. Furthermore, because $I_+$ commutes with $J_3$, we may 
set 
	\be
	|\lambda_m,+;t\kt =|j,m\kt,
	\label{lambda-m=}
	\end{equation}
where $|j,m\kt$ are the well-known orthonormal angular basis vectors satisfying
	\be
	J_3|j,m\kt=m|j,m\kt,~~~~{\bf J}^2|j,m\kt=j(j+1)|j,m\kt\;,
	\label{j-m}
	\end{equation}
$j\in\{0,\pm 1/2,\pm 1,\pm 3/2,\cdots\}$ labels the total angular momentum (spin) of the particle,
and $m\in\{-j,-j+1,\cdots,j-1,j\}$ is the magnetic quantum number. Now, in view of 
Eqs.~(\ref{I+-spin}), (\ref{-+=}) and (\ref{lambda-m=}), the eigenvalues of $I_+$ are given by
	\be
	\lambda_m=\br j,m|I_+|j,m\kt=j(j+1)-m(m+1)\,.
	\label{egva+=}
	\end{equation}
The solutions of the Schr\"odinger equation for the Hamiltonian (\ref{spin-const}) that are associated
with this choice of $|\lambda_m,+;t\kt$ are the stationary solutions
	\be
	|\psi_m,+;t\kt=e^{-itH_+}|j,m\kt=e^{-ibtm}|j,m\kt\;.
	\label{psi+-spin}
	\end{equation}
Under the supersymmetry transformation, $|\psi_m,+;t\kt$, with $m<j$, are mapped to the following 
solutions of the Schr\"odinger equation for the Hamiltonian~(\ref{spin}).
	\be
	|\psi_m,-;t\kt=\sqrt{\frac{(j-m)(j+m+1)}{2[j(j+1)-m(m+1)]}}\: 
	e^{i[(m+1)\varphi(t)-F(t)]}
	e^{-i\varphi(t)J_3}e^{-i\theta(t)J_2}|j,m+1\kt\;.
	\label{solution-spin}
	\end{equation}
Note that here $m\in\{-j,-j+1,\cdots,j-2,j-1\}$ and we have made use Eq.~(\ref{psi=psi-1}) and the 
relations
	\bea
	&&y_+(t)=\br j,m|Y_+(t)|j,m\kt=bm,~~~y_-(t)=\br j,m|Y_-(t)|j,m\kt=f(t)m,\nn\\
	&&J_+|j,m\kt=\sqrt{(j-m)(j+m+1)}\,|j,m+1\kt.\nn
	\eea
 
Next, consider the special case of the Hamiltonians~(\ref{spin}) obtained by choosing 
$\theta$=constant and $\varphi=\omega t$ for some $\omega\in\R^+$, namely
	\bea
	H_-(t)&=&br(t)\left\{f_1(t)[\cos(\omega t)J_1+\sin(\omega t)J_2]+f_2(t)J_3\right\}\;,
	\label{special-spin}\\
	r(t)&:=&b^{-1}\sqrt{4f(t)^2+[\omega-4f(t)]\omega\sin^2\theta},\nn\\
	f_1(t)&:=&\sin\theta\:[2f(t)-\omega]/[br(t)],~~~f_2(t):=2\cos\theta f(t)/[br(t)].\nn
	\eea
These correspond to the dipole Hamiltonians for which the direction of the magnetic field
precesses about the $z$-axis and its magnitude is an arbitrary function of time. The case of the
constant magnitude is obtained by setting $f$=constant. This is the well-known case of a spin in
a precessing magnetic field originally studied in Ref.~\cite{rabi}. For a more recent treatment see
\cite{bohm-qm,nova}.

In Section~4, we restricted ourselves to the study of the quadratic invariants (the invariants that are 
second order differential operators). This restriction determined the expression for the operator
$Y_-(t)$. A choice of $Y_-(t)$ which includes cubic or higher powers of $p$ would lead to the
fermionic partner Hamiltonians that cannot be expressed as a second order differential operator.
An analogue of this restriction for the systems considered in this section is the condition
that $Y_-(t)$ should belong to the Lie algebra  $su(2)=so(3)$. Unlike the case of the harmonic 
oscillators, a violation of this condition does not lead to any serious problem for the spin systems.
For example, if we take
	\be
	Y_-(t)=f(t)J_3+g(t)J_3^2\;,
	\label{quad}
	\end{equation}
but keep the same choice for $W_-(t)$, i.e., (\ref{su2}), we are led to a class of exactly
solvable fermionic partner Hamiltonians of the form
	\be
	H'_-(t)=H_-(t)+\tilde H_-(t)\;,
	\label{quad-H}
	\end{equation}
where $H_-(t)$ is given by Eq.~(\ref{spin}) and $\tilde H_-(t)$ is a general quadratic Stark
Hamiltonian describing the quadrupole interaction of a spinning particle with the magnetic field
\cite{mead}. A straightforward calculation yields
	\bea
	\tilde H_-(t)&=&g(t) W_-(t)J_3^2W_-(t)^\dagger =g(t)[W_-(t)J_3W_-(t)^\dagger]^2=
	g(t)\left[\sum_{i=1}^3\tilde R^i(t)J_i\right]^2,
	\label{quad-H=}\\
	\tilde R^1(t)&:=&\sin\theta(t)\cos\varphi(t),~~~~\tilde R^2(t):=\sin\theta(t)\sin\varphi(t),~~~~
	\tilde R^3(t):=\cos\theta(t).\nn
	\eea

The Hamiltonian~(\ref{quad-H=}) belongs to the class of quadrupole Hamiltonians 
	\be
	\tilde H(t)=\sum_{i,j=1}^3 Q_{ij}(t)J_iJ_j\;,
	\label{general-quad}
	\end{equation}
whose algebraic and geometric structure has been studied in Refs.~\cite{q1,q2}. In particular, up to
trivial addition of a multiple of identity, any quadrupole Hamiltonian may be written in the form
$\tilde H(t)=\sum_{\alpha=0}^4 \rho^\alpha e_\alpha$ where $\rho^\alpha$ are real parameters and
	\bea
	e_0&:=&J_3^2-{\bf J}^2/3,~~~e_1:=(J_1J_3+J_3J_1)/\sqrt 3,~~~
	e_2:=(J_2J_3+J_3J_2)/\sqrt 3,\nn\\
	e_3&:=&(J_1^2-J_2^2)/\sqrt 3,~~~e_4:=(J_1J_2+J_2J_1)/\sqrt 3.\nn
	\eea
Furthermore, the commutators $[e_\alpha,e_\beta]=:T_{\alpha,\beta}$ generate the group $Spin(5)=
Sp(2)$ that acts on the set of all quadrupole Hamiltonians, \cite{q1}.\footnote{The quadrupole
Hamiltonians~(\ref{general-quad}) also arise in the study of the adiabatic evolution of a complex
scalar field in a Bianchi type IX background spacetime \cite{tjp-2000}.}

These observations suggest that one may construct supersymmetric dynamical invariants whose 
bosonic and fermionic components are linear combinations of the generators $T_{\alpha\beta}$, i.e., 
they belong to the Lie algebra of $Spin(5)$, i.e., $so(5)=sp(2)$. This in turn implies that they may be obtained from
constant elements of $so(5)$ by $SO(5)$ rotations, \cite{q1}.

Next, observe that both $e_0$ and $e_4$ commute with $J_3$. Therefore, in our construction of
the partner Hamiltonians for the constant dipole Hamiltonian~(\ref{spin-const}), we may take
$Y_-(t)= \xi(t) T_{04}$, where $\xi$ is a real-valued function of time. Now, if we take
	\[ W_-(t)=e^{\sum_{\alpha,\beta=0}^4 R^{\alpha\beta}(t)T_{\alpha\beta}},\]
for arbitrary functions $R^{\alpha\beta}$, we obtain the most general invariant $I_-(t)$ belonging
to $so(5)$. By construction, the corresponding fermionic partner Hamiltonians $H_-(t)$ will constitute
a large class of time-dependent exactly solvable Hamiltonians belonging to the Lie algebra $so(5)$.
The explicit calculation of $H_-(t)$ requires an appropriate parameterization of the operator
$W_-$ in terms of the coordinates of its parameter space. 

\section{Discussions and Conclusion}

In this article we studied supersymmetric dynamical invariants. For a given 
time-dependent Hamiltonian $H_+(t)$, we have constructed a supersymmetric 
dynamical invariant $I(t)$ and an associated 
partner Hamiltonian $H_-(t)$ such that the bosonic part of $I(t)$ is a dynamical
invariant for $H_+(t)$ and the fermionic part of $I(t)$ is a dynamical invariant
for $H_-(t)$. We have shown how the solutions of the Schr\"odinger equation for
$H_+(t)$ may be used to obtain solutions of the Schr\"odinger equation for
$H_-(t)$. 

In order to compare our approach with those of Refs.~\cite{ba-sa,cannata}, we note
that we could construct an even supersymmetric invariant of the form (\ref{I=II}) by
requiring the supersymmetric charge $\qq$ to be a dynamical invariant. It is not
difficult to show that substituting $\qq$ in the Liouville-von-Neumann equation
yields the intertwining relation 
	\be
	d \left[i\frac{d}{dt}-H_+(t)\right] = \left[i\frac{d}{dt}-H_-(t)\right] d
	\label{inter}
	\end{equation}
for the operator $d$. Note that this relation is only a sufficient condition for
$I=\{\qq,\qq^\dagger\}/2$ to be a dynamical invariant. This in turn implies that
our method is more general than that of Refs.~\cite{ba-sa,cannata}. One way to see
this is to substitute Eq.~(\ref{d=wdw}) in Eq.~(\ref{inter}). Using
Eqs.~(\ref{H-pm=}) and (\ref{Y-I=zero}), one can then reduce Eq.~(\ref{inter}) to
	\[d_0Y_+(t)=Y_-(t)d_0.\]
It is not difficult to construct operators $Y_\pm (t)$ that commute with $I_\pm (0)$
but do not satisfy this equation.

\ed